\documentstyle[12pt]{article}
\input epsf
\newcommand{\ppi}{{\overline{\phi}}}
\newcommand{\half}{{1\over2}}
\newcommand{\be}{\begin{equation}}
\newcommand{\ee}{\end{equation}}
\newcommand{\beq}{\begin{equation}}
\newcommand{\eeq}{\end{equation}}
\newcommand{\bea}{\begin{eqnarray}}
\newcommand{\eea}{\end{eqnarray}}
\newcommand{\beal}{\setcounter{letter}{1} \begin{eqnarray}}
\newcommand{\eeal}{\addtocounter{equation}{1} \end{eqnarray}}
\newcommand{\none}{\nonumber \\}
\newcommand{\req}[1]{Eq.(\ref{#1})}

\newcommand{\g}{\overline{g}}
\newcommand{\gbar}{\overline{g}}
\newcommand{\phibar}{\overline{\phi}}
\newcommand{\V}{\overline{V}}

\newcommand{\Jbar}{\overline{J}}
\newcommand{\W}{\overline{W}}
\newcommand{\Z}{\overline{Z}}

\newcommand{\mcal}{{\cal M}}
\newcommand{\fcal}{{\cal F}}
\newcommand{\gcal}{{\cal G}}
\newcommand{\jcal}{{\cal J}}
\newcommand{\qcal}{{\cal{Q}}}
\newcommand{\Mbar}{\overline{M}}
\newcommand{\Qbar}{\overline{Q}}
\newcommand{\qbar}{\overline{q}}
\newcommand{\lp}{l_p}

\begin{document}
\begin{center}
{\bf Hamiltonian Thermodynamics of Charged Black Holes}
\end{center}
\vspace{0.5 cm}
\begin{center}
{\sl by\\
}\vspace*{0.50cm}
{\bf A.J.M. Medved${}^\flat$   and G. Kunstatter$^\sharp$\\}
\vspace*{0.50cm}
{\sl
$\flat$ Dept. of Physics and Winnipeg Institute of
Theoretical Physics\\
 University of Manitoba, Winnipeg, Manitoba\\
Canada R3T 2N2\\
{[e-mail: joey@theory.uwinnipeg.ca]}}\\
\vspace*{0.50cm}
{\sl
$\sharp$ Dept. of Physics and Winnipeg Institute of
Theoretical Physics\\
 University of Winnipeg, Winnipeg, Manitoba\\
Canada R3B 2E9\\
{[e-mail: gabor@theory.uwinnipeg.ca]}
}
\end{center}
\bigskip\noindent
{\large
ABSTRACT
}
\par
\noindent
We consider the most general diffeomorphism invariant
 action in 1+1 spacetime dimensions that 
 contains a metric,
dilaton and Abelian gauge field, and has at most second derivatives of the
fields. Our action contains a  topological term (linear
in the Abelian field strength) that has not been considered in previous work. 
We impose boundary conditions  appropriate
for a charged black hole confined to a region bounded by 
a surface of fixed dilaton field
 and temperature.
 By making some 
simplifying assumptions about the quantum theory, the Hamiltonian 
 partition function
is obtained. This partition function is analyzed in some detail for
the Reissner-Nordstrom black hole and for the rotating BTZ black hole. 
\newpage

\section{Introduction}\medskip
\par
The microscopic origin of black hole entropy is currently a subject of intense investigation. The Bekenstein-Hawking entropy\cite{bekenstein} of certain extremal and near 
extremal black holes has been successfully derived by counting states in
the large coupling limit of string theory\cite{stringy}. It is important to keep in mind, however, that several other,
very different, approaches have also achieved a measure of sucess\cite{carlip}
\cite{induced}\cite{loop}.
For example, Carlip\cite{carlip} has counted edge 
states in the gauge theory formulation of 2+1 gravity and obtained the 
correct entropy for the BTZ black hole\cite{BTZ}. This calculation has taken on new importance
with the realization that many of the string inspired black holes can
be related to the BTZ geometry either by looking at their near horizon geometry\cite{near}, or by using M-theory inspired duality arguments\cite{duality}.
This suggests that the correct explanation for black hole entropy might not necessarily be tied to a specific microscopic theory, nor to any specific
low energy gravity theory: it might in some sense be universal\cite{induced}. 
It
is therefore of interest to examine the statistical mechanics of black holes in a large variety of theories, in order to look for model independent features.
A particularly useful arena for such investigations is generic dilaton gravity
in two spacetime dimensions. This class of theories provides a large number of diffeomorphism invariant, solvable theories of gravity that admit black hole solutions. Moreover, there are several specific models in this class that are
of direct physical siginificance, such as spherically symmetric gravity\cite{SSG} and Jackiw-Teitelboim gravity\cite{JT}. The latter is important because its
black hole solutions  correspond to the dimensionally reduced
BTZ black hole.\cite{ortiz}. 

The study of the Hamiltonian thermodynamics of black holes in generic vacuum dilaton gravity  was started in \cite{shelemy}, generalizing a 
formalism first  applied by Louko and Whiting\cite{lw} to spherically symmetric gravity.   The purpose of the present work is
to extend the results of \cite{shelemy} to include coupling to an Abelian 
gauge field. In particular we calculate the Hamiltonian partition function
 for  a charged charged black hole confined to a ``box'' of fixed dilaton
size. Our generic results  contain as
special cases all the black holes previously analyzed\cite{others} using Louko and Whiting's  formalism,
and provides a unified treatment of a large variety of  charged 
black holes. In order to compare our results to previous work and 
check the validity of our formalism, we will examine in some detail our 
expression for the 
partition function in the case of  spherically symmetric gravity.
We will also use our results to study the Hamiltonian thermodynamics 
of  the rotating 
BTZ black hole, which, to the best of our knowledge, has not to date been
analyzed.
\par
The paper is organized as follows. In Section 2 we review generic dilaton gravity coupled to an
 Abelian gauge field. We present the most general solution as well as a description
of the thermodynamic properties of black holes in the generic theory. For completeness, we include in the action  a topological term 
involving the Abelian field strength. This term  can only 
be constructed in two spacetime dimensions and  has not been considered in
previous work.
 In Section 3, the Hamiltonian analysis of the theory is summarized, while Section 4
derives  the boundary terms that must be added to the Hamiltonian when
considering a charged black hole in a box. Section 5 presents the Hamiltonian
partition function using the results of Section 4 and examines the
resulting thermodynamics in the semi-classical, or saddle-point
approximation. In Section 6 we analyze in
detail two specific examples: spherically symmetric charged black holes
in 3+1 Einstein gravity, and the rotating BTZ black hole. Finally, Section 7
summarizes our results and discusses prospects for future work.
 
\section{Generic Dilaton Gravity with Abelian Gauge Field}
In two spacetime dimensions, the Einstein tensor vanishes identically.
In order to construct a dynamical theory of gravity with no more than
two derivatives of the metric in the action, it is necessary to
introduce
a scalar field, traditionally called the dilaton. In the past, the
dilaton was treated as essentially a lagrange multiplier, with no 
physical or  geometrical significance. In recent years, however, it
has become clear that the dilaton plays an important role. For example,
when  the
dilaton theory is derived via dimensional reduction by imposing
spherical symmetry 
in n+2 dimensional Einstein gravity, the dilaton has a geometrical
interpetation as the invariant radius of the n-sphere. More generally, the
dilaton is instrumental in determining both the symmetries and the
topology of the solutions\cite{DGK}. 
 
In the following, we consider the  most general action functional depending on the metric tensor
${\g}_{\mu\nu}$, scalar field ${\ppi}$ and Abelian gauge field
  in two spacetime
dimensions\cite{banks,DK}:
\bea
\overline{S}[\g,\ppi, A]&=&\int d^2x \sqrt{-\g} \left[{1\over2G}\left( \half
\g^{\alpha\beta}
\partial_\alpha \ppi \partial_\beta \ppi +{1\over
l^2} {\V}(\ppi) +
D(\ppi)
R(\g)\right) \right.\none
& &\left.-{1\over4} \W(\ppi) F^{\mu\nu}F_{\mu\nu}
+{\Z(\ppi)\over \sqrt{-\g}}\epsilon^{\mu\nu}F_{\mu\nu}\right].
\label{eq: action 1}
\eea
where $G$ is the dimensionless 2-d Newton constant,
$F_{\mu\nu}=\partial_\mu A_\nu-\partial_\nu A_\mu$ and $l$ is a
fundamental
constant with dimensions of length. In addition, $\V(\ppi)$, $D(\ppi)$, $\W(\ppi)$ and $\Z(\ppi)$ are arbitrary
functions of the dilaton $\ppi$. The last term in the action is
a topological term that is only possible in two spacetime
dimensions.\footnote
{GK is grateful to R. Jackiw for pointing out this possibility.}
\par
It is convenient to eliminate the kinetic term for the scalar
field. This
can be done with an invertible field redefinition providing that 
$D(\ppi)$ is a differentiable function of $\ppi$ such that $D(\ppi)\neq0$
and ${d D(\ppi)\over
d \ppi}\neq0$ for any admissable value of $\ppi$\cite{banks,domingo1}:
\bea
g_{\mu\nu} &= &\Omega^2(\ppi) \g_{\mu\nu}
\label{eq: conf reparam} \\
\phi &=& D(\ppi)
\label{eq: field redefinitions}
\eea
where
\be
\Omega^2(\ppi) = \exp \left( \half\int {d\ppi \over (dD/d\ppi)}\right)
\ee
The electromagmetic potential is left unchanged.
In terms of the new fields, the action \req{eq: action 1} takes the form:
\bea
S[g,A] &=& {1\over 2G}  \int d^2x  \sqrt{-g}\left(\phi
R(g)+{1\over
l^2}V(\phi)\right)\\
&\,&+\int d^2x \left(-{1\over 4}\sqrt{-g}W(\phi)F^{\mu\nu}F_{\mu\nu}
   +Z(\phi)\epsilon^{\mu\nu}F_{\mu\nu}\right).
\label{eq: action 2}
\eea
where $V$, $W(\phi)$ and $Z(\phi)$  are defined as:
\bea
V(\phi) &=& {{\V}(\ppi(\phi))\over \Omega^2(\ppi(\phi))}\\
W(\phi) &=& \Omega^2(\ppi(\phi) )\W(\ppi(\phi))\\
Z(\phi) &=& { \Z(\ppi(\phi))}
\eea

We henceforth consider the action only in the form \req{eq: action 2},
keeping in mind that the physical metric in general may be different
from $g_{\mu\nu}$.\footnote{It is crucial in this regard that the black hole
thermodynamics are invariant under conformal reparametrizations of the 
form \req{eq: conf reparam}}
The field equation obtained from varying $\phi$ is:
\be
R+{1\over l^2}{dV\over d\phi}-{G\over 2}
{dW(\phi)\over \phi}F^{\alpha\beta}
   F_{\alpha\beta} +{2G\over \sqrt{-g}} {dZ\over d\phi} \epsilon^{\alpha\beta}
   F_{\alpha\beta}=0
\label{eq: fielda}
\ee
while minimizing the action with respect to variations of the metric
yields:
\be
\nabla_\mu\nabla_\nu \phi- {1\over 2l^2} g_{\mu\nu} V(\phi)
 -{3\over4}G g_{\mu\nu}W(\phi)F^{\alpha\beta}F_{\alpha\beta}
  +GW(\phi)F_\mu^\gamma F_{\nu\gamma}=0
\label{eq: field2}
\ee
Finally, the equation for the Abelian gauge field is:
\be
\nabla_\mu\left(
W(\phi)F^{\mu\nu} - 2{\epsilon^{\mu\nu}\over\sqrt{-g}} Z(\phi)
\right)
\label{eq: field3}
\ee
It follows directly from the above field equations that on shell
all the fields are left invariant by Lie derivation
along the following  Killing
 vector\cite{DGK} 
\be
k^\mu = l\epsilon^{\mu\nu}\partial_\nu \phi/\sqrt{-g}
\label{eq: killing vector}
\ee
where $\epsilon^{\mu\nu}$ is the contravariant Levi-Civita  symbol:
($\epsilon^{01}=-\epsilon^{10}=1$,etc.) and  the constant $l$ has been included to ensure that the
vector components are dimensionless.
\par
The most general solution to the field equations without the
topological
term has been found in \cite{DK}. The procedure required with the
extra
term is virtually identical, so we will omit most of the details. By going to
light cone coordinates $(z_+,z_-)$ in  conformal
gauge:
\beq
ds^2= e^{2\rho} dz_+dz_-
\eeq
where $\rho(z_+,z_-)$ is an arbitrary function, one finds that \req{eq: fielda}
reduces to:
\beq
{\partial \over \partial z_\pm}\left(W(\phi) F + 2 Z(\phi)\right)
= 0
\label{eq: wzeq}
\eeq
In the above,  $F$ is a scalar  defined implicitly  by $F^{\mu\nu} =
 F E^{\mu\nu}$, where $E^{\mu\nu}=\epsilon^{\mu\nu}/\sqrt{-g}$ is the fundamental alternating tensor.
 Thus, we find that
\beq
F={1\over W(\phi)}(q-2Z(\phi))
\label{eq: F soln}
\eeq
where $q$ is a constant that corresponds to the Abelian charge. This leads
to field equations for the metric that are completely analogous to those in
\cite{DK}, with $q\to q-2Z(\phi)$. Since these field equations are algebraic
in $F$, the solutions for the metric in the present case are precisely the
same as in $\cite{DK}$, up to this replacement. In particular
\beq
e^{2\rho} = {1\over 4}(j(\phi)-2GlM-l^2 G K(\phi;q))
\eeq
 where $M$ is a constant of integration, which will be shown below
to be the ADM mass of the solution and we have defined:
\bea
j(\phi)&=&\int^\phi_0 d \tilde{\phi} V(\tilde{\phi})
\label{eq: j}\\
K(\phi;q)&=& \int^\phi_0 d\tilde{\phi} (q-2Z(\tilde{\phi}))^2/W(\tilde{\phi})
\label{eq: K}
\eea

It is most convenient to write the final solutions in manifestly static 
coordinates by exploiting the  form of the Killing vector given above.
That is, we can choose  the spatial coordiate
to be proportional to the dilaton field:
\be
\phi = x/l
\ee
In these coordinates, the metric depends only on $x$:
\be
ds^2= -(j(\phi)-2GlM-l^2 G K(\phi;q)) dt^2
    +(j(\phi)-2GlM-l^2 G K(\phi;q))^{-1} dx^2
\label{eq: metric solution}
\ee
\par
From the above solution it is easy to see that the norm of the Killing
vector is
\be
|k|^2 = -l^2 |\nabla \phi |^2 = (j(\phi)-2GlM-l^2 G K(\phi;q)).
\label{eq: killing vector norm}
\ee
Given the above equation, it is clear that the general solution 
therefore has an
apparent horizon at the surface $\phi= \phi_0 = constant$ for $\phi_0$
given by
\be
f(\phi_0)=0
\label{eq: horizon}
\ee
where we have defined
\beq
f(\phi;M,q):=  (j(\phi)-2GlM-l^2 G K(\phi;q))
\label{eq: defn f}
\eeq
The global form of the solution, and in particular the number of
horizons, depends on the specific forms of the function $j(\phi)$ and
$K(\phi;Q)$.
\par
Before describing the Hamiltonian analysis of the theory, we 
review briefly the thermodynamic properties of the
solutions. Specifically, we assume that $\phi_0$ is the value of the 
dilaton field at an exterior, bifurcative horizon. With the 
Killing vector
normalized as in \req{eq: killing vector norm}, a straightforward
calculation reveals that the surface gravity at the horizon, defined
by:
\beq
\kappa^2 := {-\half \nabla^\mu k^\nu\nabla_\mu k_\nu}|_{\phi_0}
\label{eq: kappa squared}
\eeq
is
\bea
\kappa
&=& { f'(\phi_0)\over 2l}\nonumber\\
&=& {V(\phi_0)\over 2l} -
    {l (q-2Z(\phi_0))^2 G\over 2W(\phi_0)}
\label{eq: surface gravity}
\eea
where the prime denotes differentiation with respect to $\phi$.
\par
The Hawking temperature of the horizon can be calculated 
by analytically continuing the solution exterior to the horizon
to Euclidean time,  imposing periodicity
in the imaginary time direction and requiring the resulting solution
to be regular at the horizon. Although this is a 
standard calculation, we summarize it briefly, since it plays
an important role in determining the boundary conditions required for
the subsequent Hamiltonian analysis. 
\par
The Euclidean form of \req{eq: metric solution} is:
\beq
ds^2_E= f(\phi;M,q)dt^2_E + {1\over f(\phi;M,q)}dx^2
\label{eq: euclidean solution}
\eeq
We wish to find a coordinate transformation that puts the metric in
the form:
\beq
ds^2_E= R^2 d\theta^2 + H(R) dR^2
\label{eq: Euclidean disc}
\eeq
where $|k^2|=0$ at $R=0$ and 
  $\theta$ is an angular coordinate with period $2\pi$. This can
be accomplished by defining:
\bea
\theta &=& {t_E\over a}\\
R^2&=& a^2 f(\phi)
\eea
so that
\beq
H(R)= {4 l^2\over a^2 [f'(\phi)]^2}
\eeq
 The metric \req{eq: Euclidean disc} will be regular at
$R=0$ providing that $H(0)=1$, which  requires 
\beq
a= {2l\over f'(\phi_0;M,q)}={1\over \kappa}
\eeq
and fixes the periodicity of the Euclidean time coordinate to be
\beq
2\pi a = {4\pi l\over  f'(\phi_0;M,q)}= {2\pi\over \kappa}
\label{eq: beta}
\eeq
The Hawking temperature is then
\beq
T_H = {1\over 2\pi a}=  {f'(\phi_0;M,q)\over 4\pi l}
\label{eq: Hawking temperature}
\eeq
\par
As discussed in \cite{DGK}, the expression for the black hole entropy
can most easily be derived by demanding that the first law of
thermodynamics be satisfied with respect to infinitesmal variations
of the mass and charge of the black hole. In particular, if we vary the parameters $M$ and $q$ of the solution while staying on the event horizon, $f=0$,
we get the condition on the corresponding variation of $\phi_0$ at the horizon:
\bea
0&=&{\partial f\over \partial
\phi_0}\delta \phi_0  + {\partial f\over \partial M}
\delta M +{\partial f\over \partial q}\delta q \nonumber\\
&=& \left(V(\phi_0)-{l^2G(q-2Z(\phi_0))^2\over W(\phi_0)}\right)
\delta \phi_0 - 2Gl^2 \delta M - {\cal P}(\phi_0,q)\delta q
\label{eq: horizon variation}
\eea
where 
\beq
{\cal P}(\phi_0,q) = \int^{\phi_0} d\phi 
 { (q- 2 Z(\phi))\over W(\phi)}
\label{eq: generalized force}
\eeq
This yields the first law of black hole thermodynamics:
\beq
\delta M = T_H \delta S_{BH} - {\cal P} \delta q
\label{eq: first law}
\eeq
where we have defined the Bekenstein-Hawking entropy:
\beq
S_{BH}(M,q)= {2\pi\over G} \phi_0(M,q)
\label{eq: BH entropy}
\eeq
where $\phi_0(M,q)$ is obtained by solving \req{eq: horizon}.
\req{eq: first law} also shows that ${\cal P}$ is
The generalized force  associated with the charge $q$.

The expression \req{eq: BH entropy} 
for the entropy  can also obtained using Wald's
formalism\cite{wald}. It is important to keep
in mind that the thermodynamic properties defined above are {\bf not} affected
by conformal reparametrizations of the form \req{eq: field redefinitions}.

\section{Hamiltonian Analysis}
The Hamiltonian analysis for generic dilaton gravity has been presented in 
many works. Here we summarize the results, using the notation and conventions
of \cite{DK}. We start by decomposing the metric as follows:
\be
ds^2=e^{2\rho}\left[-{u}^2dt^2+\left(dx+{v}dt\right)
^2\right]
.\label{eq:adm}
\ee
where $x$ is a local coordinate for the spatial section $\Sigma$ and $\rho$,
${u}$ and ${v}$ are
functions of spacetime coordinates $(x,t)$. For convenience we work with the form of the
action in \req{eq: action 2}. In terms of the
 parametrization \req{eq:adm},
the action \req{eq: action 2} takes the form (up to surface terms):
\bea
I &=&\int dt\int^{\sigma_+}_{\sigma_-} dx[ {1\over G} \left(
{\dot{\phi}\over {u}}
({v}\rho' + {v}' - \dot{\rho}) + {\phi'\over {u}} (  {u}{u}' - {v}{v}' + {v}
\dot{\rho}
 + {u}^2\rho' - {v}^2\rho')\right.\nonumber\\
&\, &+\left.\half{u} e^{2\rho} {V(\phi)\over l^2}\right)
 + {e^{-2\rho}\over2\mu}
   W(\phi)(\dot{A}_1-A'_0)^2 +2Z(\phi)(\dot{A}_1-A'_0)]
\label{eq: action 3}
\eea
In the above, dots and primes denote differentiation with respect to time and
space, respectively, while $\sigma_+$ and $\sigma_-$ are the outer and inner spatial
boundaries.
The canonical momenta for the fields
 $\{\phi, \rho\}$ are:
\bea
\Pi_\phi & = &{1\over G{u}} ({v}\rho'+{v}'-\dot{\rho})
\label{eq: Pi phi}\\
\Pi_\rho &=& {1\over G{u}} (-\dot{\phi} +{v}\phi')
\label{eq: pirho}\\
\Pi_{A_1}&=& {e^{-2\rho}\over {u}} W(\phi)(\dot{A}_1-A'_0) +2Z(\phi)\\
\Pi_\mu&=&\Pi_v=\Pi_{A_0}=0
\eea
As expected, the momenta conjugate to ${u}$,$v$ and $A_0$ vanish
because these fields play the
role of Lagrange
multipliers that are needed to enforce the first class constraints associated
with diffeomorphism and gauge
of the classical action. A straightforward calculation
leads to the canonical
Hamiltonian (up to surface terms which will be discussed below):
\be
H_c= \int dx\left({v}\fcal+ {{u}\over 2G} \gcal + A_0 \jcal \right)
\label{eq: canonical hamiltonian}
\ee
where 
\bea
{\fcal}&=& \rho'\Pi_\rho +\phi'\Pi_\phi-\Pi'_\rho \sim 0
\label{eq: f constraint}\\
\gcal&=& 2\phi''-2\phi'\rho' -2G^2\Pi_\phi \Pi_\rho - e^{2\rho} {V(\phi)\over
l^2}\nonumber\\
 &\,& +{Ge^{2\rho}\over W(\phi)}[\Pi_{A_1}-2Z(\phi)]^2\sim 0
\label{eq: g constraint}\\
\jcal &=& -\Pi'_{A_1}
\label{eq: Gauss law}
\eea
are secondary constraints. Note that $\fcal$ annd $\gcal$ generate spatial and
temporal diffeomorphisms, while $\jcal$ is the Gauss law constraint that 
generates Abelian gauge transformations.
\par
The general solution presented in the previous section suggests that 
there are  two independent, diffeomorphism invariant
 physical observables, namely the mass of the black hole and its
Abelian charge. These observables can easily be expressed in terms of
the phase space variables. In particular, define:
\beq
{\cal Q}=\Pi_{A_1}
\eeq
Clearly, $\cal{Q}$ commutes with all three constraints and hence the
entire canonical Hamiltonian, and the
Gauss law constraint implies that ${\cal Q}=q$ is constant
on the constraint surface. The constant mode $q$ of ${\cal Q}$ is therefore a
physical observable and corresponds precisely to the Abelian charge in
the solution \req{eq: F soln}. Similarly, we can define the mass
observable:
\beq
\mcal = {l\over 2G}\left(e^{-2\rho}(G^2\Pi_\rho^2-(\phi')^2)
  +{j(\phi)\over l} - G K(\phi,\qcal)\right)
\label{eq: mass observable}
\eeq
where 
\be
K(\phi,\qcal) := \int^\phi d\tilde{\phi} {(\qcal - 2Z(\tilde{\phi}))^2\over 
W(\tilde{\phi)}}
\ee
Once again it is possible to show that $\cal M$ commutes with the
constraints and is spatially constant on the constraint surface. In
particular, we find that:
\be
{\partial \mcal\over \partial x} =
-l e^{-2\rho}\left(G\Pi_\rho \fcal +{1\over 2G} \phi' \gcal
-e^{2\rho} {\cal P}(\phi,\qcal)\jcal\right)
\label{eq: mass constraint}
\ee
where 
\be
{\cal P}(\phi,\qcal) = \int d\phi {(\Pi_{A_1} -2Z(\phi))\over W(\phi)}
\ee
 The
constant mode of $\cal M$ is the mass parameter appearing in the
solution \req{eq: metric solution}.

It is useful to note that both $\cal M$ and $\cal Q$ can be written as
coordinate invariant scalars in terms of the dilaton and  the Abelian
field strength as follows:
\bea
\mcal &=& {1\over 2Gl} \left(|k|^2 + j(\phi) - l^2 G K(\phi,\qcal)\right)
\label{eq: mass2}\\
{\cal Q} &=& 2Z(\phi) + \left(-{W(\phi)\over2}
  F^{\mu\nu}F_{\mu\nu}\right)^{1\over 2}
\label{eq: charge2}
\eea
For completeness we also write down the explicit expressions for the
momenta canonically conjugate to the mass and charge observables. They
are, respectively,
\bea
\Pi_\mcal &=& -{G\over l} \int dx {e^{2\rho} \Pi_\rho\over
  (G\Pi_\rho)^2 - (\phi')^2}\\
\Pi_{\cal Q} &=& -\int dx
\left(A_1 + { G e^{2\rho} \Pi_\rho {\cal P}(\phi\qcal)\over 
 (G\Pi_\rho)^2 - (\phi')^2}\right)
\eea
Although the observables $\cal M$ and $\cal Q$ are invariant under
general diffeomorphisms, their conjugates $\Pi_\mcal$ and $\Pi_{\cal Q}$ 
are only invariant with respect to diffeomorphisms that vanish on the
boundaries of the system. The Hamiltonian analysis is therefore
consistent with the results of the previous section which indicate that,
up to general diffeomorphisms, there exists only  a two parameter family of
physically distinct solutions.
\section{Boundary Terms in the Hamiltonian}
The previous Section neglected the boundary terms that must be added
to the canonical Hamiltonian in order that the variational principle
be well defined. These depend on the boundary conditions and define
the canonical energy, since the remainder of the Hamiltonian vanishes
on the constraint surface. We now derive the boundary terms for
boundary conditions corresponding to a charged black hole in a box of
fixed, constant ``radius'' (surface of constant dilaton field). For
convenience we rewrite the canonical Hamiltonian as follows:
\be
H_c = \int^{\sigma_+}_{\sigma_-} \int dx
\left( {\tilde {u}} \tilde{\gcal}
+ {\tilde v}{\fcal} +
{\tilde A}{\jcal} \right)
+ H_+ - H_-
\label{eq: canHam}
\ee
where have replaced the original Hamiltonian constraint $\gcal$
by the linear combination of constraints corresponding to the spatial
derivative of the mass observable:
\be
\tilde{\gcal} = -{\partial \mcal\over \partial x} =
l e^{-2\rho}\left(G\Pi_\rho \fcal +{1\over 2G} \phi' \gcal
-e^{2\rho} {\cal P}\jcal\right)
\label{eq: gcal}
\ee
and replace the original lagrange multipliers by:
\bea
\tilde{{u}} &=& { {u} e^{2\rho} \over l \phi'}\\
\tilde{v} &=& v-{{u} G \Pi_\rho \over \phi'}\\
\tilde{A} &=& A_0 + {{u} e^{2\rho} \over \phi'} {\cal P}
\eea
$H_+$ and $H_-$ are previously neglected boundary terms
 determined by the requirement that the surface
terms in the variation of $H_c$ 
vanish for a given set of boundary conditions.
\par
We wish to consider the 1+1
dimensional analogue of a charged black hole in a box of fixed radius. We will
therefore keep the value of the dilaton at the 
outer 
boundary $\phi_+:=\phi(\sigma_+)$
 fixed and independent
of time, as well as  the component of the metric along the world line
of the box ($g_{tt}^+:=g_{tt}(\sigma_+)$). 
Note that $\dot{\phi_+}=0$ requires that $\tilde{v}(\sigma_+)=0$ 
(cf \req{eq: pirho}). The relevant 
 boundary conditions on the vector potential are $A_1(\sigma_+) = 0$ and $A_0(\sigma_+)= A_0^+=constant$.
Give the above conditions, the boundary  variation of 
the canonical Hamiltonian \req{eq: canHam} at $\sigma_+$ will vanish if:
\be
\delta H_+(\mcal,q) =\tilde{{u}}\delta\mcal|_{\sigma_+} + \tilde{a}
    \delta \qcal|_{\sigma_+}
\label{eq: H+}
\ee
where we have used \req{eq: gcal}   and the fact that
$\jcal=-\qcal'$. Moreover, since
\bea
\tilde{{u}}_+&=&\left({g_{tt}^+\over 2G\mcal l - j(\phi_+)
    + l^2 G K(\phi_+,\qcal)}\right)^\half\\
\tilde{A}_+&=& A^+_0 + {l\over 2} \tilde{{u}}(\sigma_+)
  \left.{\partial K(\phi_+,\qcal)\over \partial \qcal}\right|_{\sigma_+}
\eea
\req{eq: H+} can be directly integrated to yield:
\be
H_+(\mcal,\qcal) ={\sqrt{-g_{tt}^+j(\phi_+)}\over lG}
  \left(1- \sqrt{1-{2Gl\mcal\over j(\phi_+)} - {l^2GK(\phi_+,\qcal)\over
   j(\phi_+)}}\right)+ A^+_0 \qcal
\ee
Note that  we have chosen the constant of integration so as to guarantee that
$H_+=0$ when $\mcal=\qcal=0$. If
  $K(\phi_+,\qcal)$ remains finite as $\phi_+\to \infty$, then 
\be
H_+(\mcal,\qcal)\to \sqrt{-g_{tt}^+\over j(\phi_+)}\mcal 
\ee
Hence, on the constraint surface, $\mcal$ is proportional to the 
ADM  mass. The value of the constant of proportionality will depend
on the boundary conditions on the metric and $\phi_+$. This will be 
discussed in more detail below.
\par
We next consider the inner boundary $\sigma_-$. Following the work of 
Louko and Whiting\cite{lw} we require our spatial slices to approach the
bifurcation point ($k^\mu=0$) of the black hole along a static slice. These 
boundary conditions are natural for the consideration of the thermodynamics
of the black hole, since the resulting spacetimes can be analytically continued
to the Euclidean spacetime described by the non-singular Gibbons-Hawking
instanton. Given the general form of the Killing vector in \req{eq: killing vector}, for a static
slice ($\dot{\phi_-}=0$),
the condition that $\sigma_-$ be a bifurcation point reduces to:
\be
\phi'(\sigma_-)=0
\ee
From the thermodynamic considerations of Section 2, it follows
that the metric on the inner boundary must approach the form:
\be
ds^2\to -R^2(dt/\tilde{a})^2 + H(R) dR^2
\ee
where $R=0$ at the bifurcation point $\sigma_-$,  
$H(0)=1$ and $2\pi \tilde{a}$ equals
 the periodicity of the Euclidean time required to make the Euclidean
solution regular at the horizon.\footnote{Recall that the time coordinate
in this Section is  normalized so that $g_{tt}^+$ is fixed. The 
parameter $\tilde{a}$ therefore differs from $a$ in Secton 2.}
 The required 
boundary conditions on the metric components in $(t,R)$ coordinates 
are therefore
\bea
e^{2\rho(\sigma_-)}&=&1\\
v(\sigma_-) &=& 0\\
{u}(\sigma_-)&=&0\\
u'(\sigma_-)&=& {1\over \tilde{a}}
\eea
Since, in terms of phase space coordinates,
\beq
|k|^2 = l^2e^{-2\rho}((G\pi_\rho)^2-\phi'^2)
\eeq
we must also impose the
condition\footnote{One might expect to conclude this from \req{eq: pirho}, but
this is not possible without further assumptions because $u=0$.}:
\beq
\pi_\rho(\sigma_-)=0
\eeq
to ensure that $|k|_-^2=0$.
\par
Finally, following Louko and Winters-Hilt\cite{others}, we choose  the boundary conditions on the U(1)
vector potential at the bifurcation point to be:
\bea
A_1(\sigma_-)&=&0\\
A_0(\sigma_-)&=&A_0^-= constant
\eea
\par
With the above boundary conditions we find:
\bea
\tilde{v}(\sigma_-) &=&0\\
\tilde{{u}}(\sigma_-)&=&{2l\over  \tilde{a} \tilde{V}(\phi_-,\qcal)}\\
\tilde{A}(\sigma_-)&=& {l^2\over  \tilde{a}\tilde{V}(\phi_-,\qcal)}
 {\partial K(\phi_-,\qcal)\over \partial \qcal}+ A_0^-
\eea
where we have defined
\beq
\tilde{V}(\phi,\qcal)= V(\phi) - Gl^2 {\partial K(\phi,\qcal)\over \partial\phi}
\eeq
The boundary value for $\tilde{{u}}$ was  obtained by
applying l'Hopital's rule and then using the constraint \req{eq: g constraint}
to eliminate $\phi''(\sigma_-)$.  $A_0^-$ and $\tilde{a}$ are considered
fixed parameters.
On the other hand, $\phi_-:=\phi(\sigma_-)$ is {\bf not} constrained, but is
a dynamical variable. In fact, an implicit equation for $\phi_-$ in terms of the
physical observables $\mcal$ and $\qcal$ can be obtained by setting  $|k|^2=0$ 
in the expression for 
$\mcal$ (i.e. \req{eq: mass2}). 
\par
With these boundary conditions there will  be no boundary terms
at $\sigma_-$ from the variation of the Hamiltonian if:
\be
\delta H_- = {2l\over \tilde{a}\tilde{V}(\phi_-,\qcal)}\delta\mcal|_{\sigma_-}
 + \left.{l^2 \over  \tilde{a} \tilde{V}(\phi_-,\qcal)} 
{\partial K(\phi_-,\qcal)\over
\partial \qcal}\right|_{\sigma_-} \delta\qcal+ A_0^-\delta\qcal
\label{eq: delta H-}
\ee
Next we use the fact that the norm of the Killing vector is constrained
to vanish at the inner boundary to obtain (via \req{eq: mass2})
\be
\delta\mcal = {1\over 2Gl} \tilde{V}(\phi_-,\qcal)\delta \phi_-
 -{l\over 2}{\partial K(\phi_-,\qcal)\over \partial \qcal}\delta \qcal
\ee
Substituting this into \req{eq: delta H-} and simplifying gives:
\be
\delta H_- = {1\over \tilde{a} G}\delta \phi_- + A_0^- \delta \qcal
\ee
which can be trivially integrated to yield:
\be
H_-(\mcal,\qcal) = {1\over \tilde{a} G} \phi_-(\mcal,\qcal) + A_0^-\qcal
\label{eq: H-}
\ee
By using \req{eq: BH entropy}
our final expression for the  canonical Hamiltonian on the constraint
surface  takes the simple form:
\be
H_c = E(M,q;\phi_+) - {1\over 2 \pi \tilde{a}} S_{B.H.}(M,q) - \gamma q
\label{eq: final Ham}
\ee
where 
\be
E(M,q;\phi_+) = {\sqrt{-g_{tt}^+ j(\phi_+)}\over Gl}
\left(1-\sqrt{1- {2GMl\over j(\phi_+)}-{l^2GK(\phi_+,q)\over j(\phi_+)}}
\right)
\label{eq: quasilocal energy}
\ee
is the quasilocal energy and $\gamma\equiv A_0^--A_0^+$. We have also 
used the fact that on the constraint surface
$\mcal=M$ and $\qcal=q$, where $M$ and $q$ correspond to the physical mass
and charge appearing in the general solution \req{eq: metric solution}.
\par

In addition to the 
dynamical variables $M$ and $q$, the canonical Hamiltonian appears to depend
on four fixed  external 
parameters, $g_{tt}^+$, $\phi_+$, $\tilde{a}$  and $\gamma$. $\phi_+$ plays the role of
 the effective box size, while $\gamma$ is analogous to a 
chemical potential. $g_{tt}^+$ and $\tilde{a}$ on the other hand 
must be fixed by imposing further boundary conditions. In particular, the metric $g_{tt}^+$  is related to the choice of time coordinate along the boundary.
This is normally chosen to equal the proper time as measured
with respect to a given physical metric. In vacuum dilaton gravity,
the choice of physical metric is subtle since one can always do conformal reparametrizations involving
the dilaton. One must therefore define the ``physical metric'' to be the one
which determines the geodesics of massive test particles. This cannot
be determined {\it a priori}, but must ultimately be settled by experiment. 
Since $g_{\mu\nu}$ was arrived at from the original metric 
$\overline{g}_{\mu\nu}$ by a conformal reparametrization designed to make
the action simpler(cf \req{eq: conf reparam}), the physical metric might
be
$\overline{g} = \Omega^{-2}g$ as given in \req{eq: conf reparam}. In this
case,   we would set  $\overline{g}_{tt}^+=-1$  so that 
\beq
g_{tt}^+= - \Omega^2(\phi_+)
\eeq
 On the other hand, the metric
\beq
\tilde{g}_{\mu\nu} = {g_{\mu\nu}\over j(\phi)}
\label{eq: tilde g}
\eeq
has the desirable property that it approaches the Minkowski metric 
as $\phi\to \infty$, so one might be tempted to define this as the physical
metric\footnote{It is interesting to note that for spherically symmetric
gravity in n+2 dimensions\cite{shelemy}, $\tilde{g}$ and $\overline{g}$ are
equal, and coincide with the projection onto two spacetime dimensions of the
$n+2$ dimensional physical metric.} Thus, if we set $\tilde{g}_{tt}^+=-1$,
then 
\beq
g_{tt}^+= - {j(\phi_+)}
\eeq 
With this choice of normalization, the quasilocal energy $E\to M$ as $\phi_+\to \infty$.  For now we will consider the most general case and
write 
\beq
g_{\mu\nu} = h(\phi)g^{phys}_{\mu\nu}
\label{eq: physical metric}
\eeq
 where $h(\phi)$ is an arbirary function of $\phi$ that must ultimately
be determined experimentally.
If  $g^{phys}_{tt}=-1$, then
\beq
g_{tt}^+ =- {h(\phi_+)}
\label{eq: alpha}
\eeq
\par

The constant $\tilde{a}$ must be fixed by thermodynamic
considerations\cite{lw}. We have already shown that  $2\pi \tilde{a}$ must be
equal to the period of the corresponding Euclidean time in order for the 
Euclideanized solution to be regular at the horizon. In
the Euclidean 
formulation of black hole thermodynamics, the inverse
 temperature $\beta$ at the boundary
of the system is
\beq
\beta = \sqrt{-g_{tt}^{phys}(\sigma_+)} 2\pi \tilde{a}
\label{eq: beta1}
\eeq
Thus, if as discussed above $g_{tt}^{phys}(\sigma_+)=-1$, we find that
\beq
\tilde{a}={\beta\over 2\pi}
\eeq
\par
The final form of the canonical Hamiltonian is therefore:
\beq
H_c =  E(M,q,\phi_+) - \beta^{-1} S_{B.H.}(M,q) - \gamma q
\label{eq: final Ham2}
\eeq
where
\be
E(M,q;\phi_+) = {\sqrt{h(\phi_+) j(\phi_+)}\over Gl}
\left(1-\sqrt{1- {2GMl\over j(\phi_+)}-{l^2G^2K(\phi_+,q)\over j(\phi_+)}}
\right)
\label{eq: final energy}
\ee
We will now examine the properties of the resulting  partition function.

\section{Hamiltonian Partition Function}

The quantum partition function of interest is formally defined as:
\beq
Z[\beta,\phi_+,\gamma]= Tr[\exp(-\beta \hat{H})]
\label{eq: formal partition}
\eeq
where the trace is over all physical states and $\beta$ corresponds to the
(fixed) temperature at the boundary of the system. 
This trace is most easily
expressed in term of the eigenstates $|M,Q>$ of the mass and charge operators:
\bea
Z(\beta,\phi_+,\gamma)&=&\int dM \int dQ \, \mu(M,Q)
   <M,Q|e^{-\beta\hat{H}}|M,Q>
\label{eq: partition 2}
\eea
In the above, $\mu(M,Q)$
is an as yet unknown measure on the space of observables. In principle
both the spectrum of observables and the measure should  be derivable
from a rigorous quantization procedure.  Exact
eigenstates of the mass and charge operators can be found within a
Dirac quantization scheme in which the constraints annihilate physical
states. This procedure yields a continuous and unbounded mass spectrum for
Lorentzian black holes\footnote{Interestingly, a discrete spectrum has
  been obtained via this procedure for Euclidean black holes in
  generic dilaton gravity}.  In this case the inner product 
$<M,Q|M,Q>$ and the choice of  measure is problematic. Following
Louko and Whiting\cite{lw}, we will make the
simplest, physically reasonable assumptions about the measure and the allowed values
of $M$ and $Q$. A more rigorous derivation of the
measure will be addressed in future work. First of all, we
restrict the ADM mass $M$ to be positive. Secondly, we
allow only those value of $M$ and $Q$  for which at least one bifurcative
horizon exists where  $f(\phi)$ has a simple zero (i.e. no extremal black
holes or naked singularities).
 Finally, we require the value of the dilaton at the horizon to be
less than its value at the boundary of the system (ie the box must lie
outside the horizon) so that equilibrium is in fact possible. With
these assumptions the space of allowed values for the observables is
finite. This will be made explicit for specific examples in the next section.
 
As in \cite{lw} (see also \cite{bose2})
we assume that
\beq
\mu(M,Q)<M,Q|M,Q>= {1\over {\cal {V}}}
\label{eq: measure}
\eeq
where ${\cal V}$ is the volume of the allowed space of observables.
The final
 expression for the partition function is therefore:
\beq
Z(\beta,\phi_+,\gamma)={\cal V}^{-1}\int_{\cal V} dM dq e^{S_{BH}(M,q)}
   e^{-\beta \left(E(M,q,\phi_+) - \gamma q\right)}
\label{eq: partition 3}
\eeq
Note that the Bekenstein-Hawking entropy enters the partition function
as the logarithm of an apparent degeneracy of the physical mass and
charge eigenstates. Moreover, $q$ is thermodynamically analoguous to
particle number, while $\gamma$ plays the role of a chemical potential.

\par
The above expression, can in principle be integrated to yield the
partition function describing the thermodynamics  charged black holes in
a box for any particular
dilaton gravity theory. We will now show that it gives  the
correct classical thermodynamic behaviour in the saddle-point
approximation. In this approximation, the choice of measure
is irrelevant except in the unlikely event that it is exponential in
the observables. Thus, we have
\beq
Z(\beta,\phi_+,\gamma)\approx e^{-I(\Mbar,\qbar,\beta,\phi_+,\gamma)}
\label{eq: saddle Z}
\eeq
where we have defined:
\beq
I(M,q,\beta,\phi_+,\gamma) = \beta(E(M,q,\phi_+) - \gamma q) -S_{BH}(M,q)
\label{eq: def i}
\eeq
and $\Mbar$ and $\qbar$ are the values of the mass and charge at the
minimum of $I$ (if one exists). The equation obtained by extremizing with
respect to $M$ is:
\bea
0=\left.{\partial I\over \partial M}\right|_{\Mbar,\qbar}&=&
   \left.\left(
\beta{\partial E\over \partial M} - {\partial S_{BH} \over \partial M}
    \right)\right|_{\Mbar,\qbar}\nonumber\\    
   &=&\left.\left(\beta \sqrt{h(\phi_+)}{M\over \sqrt{f(\phi_+,M,q)}}
  -\beta_H(M,q) \right)  \right|_{\Mbar,\qbar}
\label{eq: mean M}
\eea
where $\beta_H = 1/T_H = 4\pi l / f'(\phi_-,M,q)$. This implies that, semi-classically, the temperature at the boundary is the Hawking temperature $T_H$
red-shifted with respect to the physical metric $g_{phys}$:
\beq
\beta = \sqrt{{f(\phi_+,\Mbar,\qbar)\over h(\phi_+)}} \beta_H(\Mbar,\qbar)
\label{eq: red shift}
\eeq
Variation with respect to $q$ gives:
\bea
0={\partial I\over \partial Q}&=&
  \beta \left( {\partial E\over \partial q} - \gamma\right)
   -{\partial S_{BH}\over \partial q}\nonumber\\
  &=& {\beta_H l\over 2} \left(
   {\partial K(\phi_+,q)\over \partial q}-
     {\partial K(\phi_-,q)\over \partial q}\right)
 -\beta \gamma
\label{eq: mean q}
\eea
where as $\phi_-=\phi_-(M,q)$ as determined by \req{eq: horizon}. This
then yields an expression for the chemical potential in terms of the
 $\Mbar$, $\qbar$ and the inverse temperature, $\beta$:
\beq
\gamma = {l\over 2}\left.{\beta_H(\Mbar,\qbar)\over \beta}
  \left(
   {\partial K(\phi_+,q)\over \partial q}-
     {\partial K(\phi_-,q)\over \partial q}\right)\right|_{\Mbar,\qbar}
\label{eq: gamma}
\eeq
Using \req{eq: saddle Z}  we can evaluate the mean energy, mean charge and
entropy of the system:
\bea
<E> &=& - \left.{\partial \ln(Z) \over \beta}\right|_{\Mbar,\qbar}
   +{\gamma\over \beta}\left. {\partial \ln(Z)\over \partial \gamma}\right|_{\Mbar,\qbar}
 \nonumber\\
  &\approx& E(\Mbar,\qbar,\phi_+)
\label{eq: mean energy}\\
<q>&=& \beta^{-1}
\left.{\partial \ln(Z)\over \partial \gamma}\right|_{\Mbar,\qbar}
   \approx \qbar
\label{eq: mean charge}\\
S&=& \left(1-\beta{\partial \,\,\over\partial\beta} \right) \ln(Z) =
   S_{BH}(\Mbar,\qbar)
\label{eq: grav entropy}
\eea
A straightforward calculation verifies that the above expressions 
for the mean energy, charge and entropy automatically
obey the generalized first law
\bea
\delta<E> &=& {\partial E\over \partial M}\delta \Mbar
  +{\partial E\over \partial q}\delta \qbar 
+  {\partial E\over \partial \phi_+}\delta \phi_+
\nonumber\\
   &=&  \beta^{-1}\delta S_{BH} + \gamma\delta <q>
      -{\cal W} \delta\phi_+
\label{eq: first law2}
\eea
where
\beq
{\cal W} := -\left.{\partial E(M,q,\phi_+)\over \partial \phi_+}
   \right|_{\Mbar,\qbar}
\label{eq: surface pressure}
\eeq 
is a generalized surface pressure: it is the rate of change of quasilocal
energy with ``box size''. The final expression in \req{eq: first law2} was 
obtained by using the mean field equations \req{eq: mean M} and \req{eq: mean q} to express $\partial E/\partial M$ and $\partial E/\partial q$ in terms
of the derivatives of $S_{BH}$ with respect to $M$ and $E$.
\section{Examples}
\subsection{Spherically Symmetric Gravity}
The action for four dimensional Einsten-Maxwell theory is:
\be
I^{(4)}= {1\over 16 \pi G^{(4)}}\int d^2x \sqrt{-g^{(4)}}
 \left( R(g^{(4)}) -
       F^{AB}F_{AB}\right)
\ee
where the indices $A,B=0,1,2,3$ and $G^{(4)}$ is the four dimensional
Newtonion constant. For convenience we have
rescaled the vector potential by a multiple
of the 4-D Planck length $l_p=\sqrt{G^{(4)}}$ in order to make it
dimensionless. We impose 
spherical symmetry via the {\it ansatz} 
\bea
ds^2_{(4)} &=& \gbar_{\mu\nu} dx^\mu dx^\nu + {l^2\phibar^2\over2}
(d\theta^2 + \sin^2\theta
              d\phi^2)\\
A_Adx^A&=& A_\mu(x^\mu) dx^\mu
\eea
with $\mu, \nu = 0,1$. Note that  $l^2\phibar^2/2=r^2$ where
$r$ is the usual radial coordinate. Here, $\phibar$ and $r$ are taken
to be functions of the coordinates $x^\mu$. $l$ is an arbitrary constant
of dimension length, and without loss of generality we take it to 
equal the four dimensional Planck length $\lp$.
 After integrating over the angular variables,
 the reduced action takes the form of a dilaton 
gravity theory in two spacetime dimensions:
\be
I = {1\over 2}\int d^2x \sqrt{-\gbar}
\left({\phibar^2\over 4} R(\gbar) + {1\over \lp^2} +
{1\over 2} |\nabla \phibar|^2\right)
   -{1\over 4} \int d^2x \sqrt{-g} {\phibar^2\over 2} F^{\mu\nu}F_{\mu\nu}
\ee
This is the same form as \req{eq: action 1} with
$D(\phibar) = \phibar^2/4$, $W(\phibar)= \phibar^2/2$ and $G=1$. We now make the field redefinitions
\bea
g_{\mu\nu} &=& \Omega^2(\phibar) \g_{\mu\nu}\\
\phi&=& {1\over4 } \phibar^2
\eea
with 
\be
\Omega^2(\phibar) =\exp\half\int {d\phibar\over (dD(\phibar)/d\phibar))}
  = {\phibar\over \sqrt{2}}
\ee
In the above, the integration constant was chosen to be $1/\sqrt{2}$. As
we will see in the subsequent analysis, this choice
guarantees that the physical metric $\overline g$ has the correct asymptotic behaviour.  The final action is then  of the same form as 
\req{eq: action 2}, with $G=1$, $Z(\phi)=0$, $V(\phi) = 1/(\sqrt{2\phi})$
and $W(\phi) = (2\phi)^{3/2}$. 
The solution  for the   metric $g$ is therefore:
\be
ds^2 =  -\left(\sqrt{2\phi} - 2M\lp + {Q^2\lp^2\over \sqrt{2\phi}}\right) dt^2
    + \left(\sqrt{2\phi} - 2M\lp + {Q^2\lp^2\over \sqrt{2\phi}}\right)^{-1}dx^2
\label{eq: ssg solution 1}
\ee
In terms of the radial coordinate $r=\l\sqrt{2\phi}$, the solution takes the
form:
\be
ds^2 = {r\over \lp}\left(
- \left(1 - {2M\lp^2\over r} + {Q^2\lp^4\over r^2}\right) dt^2
    +  \left(1 - {2M\lp^2\over r} + {Q^2\lp^4\over r^2}\right)^{-1}dr^2\right)
\label{eq: ssg solution 2}
\ee
The physical metric is therefore the usual
Reissner-Nordstrom solution:
\be
d\overline{s}^2 = 
- \left(1 - {2M\lp^2\over r} + {Q^2\lp^4\over r^2}\right) dt^2
    +  \left(1 - {2M\lp^2\over r} + {Q^2\lp^4\over r^2}\right)^{-1}dr^2
\label{eq: ssg solution 3}
\ee
The solution for the scalar $F$ \req{eq: F soln} is:
\be
F= {Q\lp^3\over r^3}
\ee
so that the electromagnetic field strength is 
\be
F_{01} = {Q \lp^2\over r^2}
\ee
as expected.
 The solution \req{eq: ssg solution 2}  has event horizons at:
\be
r_{o,i} = \lp^2(M \pm \sqrt{M^2-Q^2})
\label{eq: ssg horizons}
\ee
where $r_o$  and $r_i$ denotes the outer and inner  horizon, respectively.
\par 
From the formulae \req{eq: Hawking temperature} and \req{eq: BH entropy}.
we can calculate the Hawking temperature associated with the outer horizon:
\bea
T_H&=&{1\over 4\pi r_o} - {\lp^4Q^2\over 4\pi r_o^3}\\
   &=& {\sqrt{M^2-Q^2}\over 2\pi \lp^2(M+\sqrt{M^2-Q^2})^2}
\label{eq: ssg Hawking temp}
\eea
and the associated Bekenstein-Hawking entropy:
\be
S_{BH}= {\pi r_o^2\over \lp^2}
\ee
which is one quarter the area of the outer horizon as required.
\par
We know discuss the thermodynamics implied by the partition function
\req{eq: partition 3}.
Since the physical metric is $\gbar$ we must choose
 $h(\phi) = \Omega^2(\phi) = r/\lp$.
The thermodynamic energy \req{eq: final energy} in the semi-classical
approximaton  is:
\be
E(\Mbar,\Qbar, r_+) = {r_+\over \lp^2}
  \left(1-\sqrt{1-{2\Mbar \lp^2\over r_+} + {\Qbar^2\lp^4 \over r_+^2}} \right)
\ee
which matches the expression for the quasilocal energy obtained
in previous work\cite{brown}. 
\par
Using \req{eq: gamma}, the semi-classical 
chemical potential for charged black holes is:
\be
\gamma = {\lp^2\Qbar \over r_o}
   \sqrt{r_+\over \lp}   
{1-r_o/r_+\over \sqrt{{r_+\over \lp}-2\Mbar\lp
    +{\Qbar^2\lp^3/r_+}}}
\ee
This expression can be simplified  significantly by expressing $\Mbar$ in terms
of the radius $r_1$ of the outer horizon:
\be
\Mbar= {r_o\over 2\lp^2} + {\Qbar^2 \lp^2\over 2 r_o}
\ee
which yields
\be
\gamma = {\lp^2\Qbar\over r_o} {\sqrt{1-r_o/r_+}\over \sqrt{1-{\Qbar^2 \lp^4 \over
   r_or_+}}}
\label{eq: ssg gamma}
\ee
Note that as the box size, $r_+$ goes to infinity, $\gamma$ approaches the
usual expression for the electrostatic potential at a distance $r_o$ from 
a charge $\Qbar$.
\par
\begin{figure}[hbt]
\begin{center}
\leavevmode
\epsfxsize=16 cm
\epsfbox{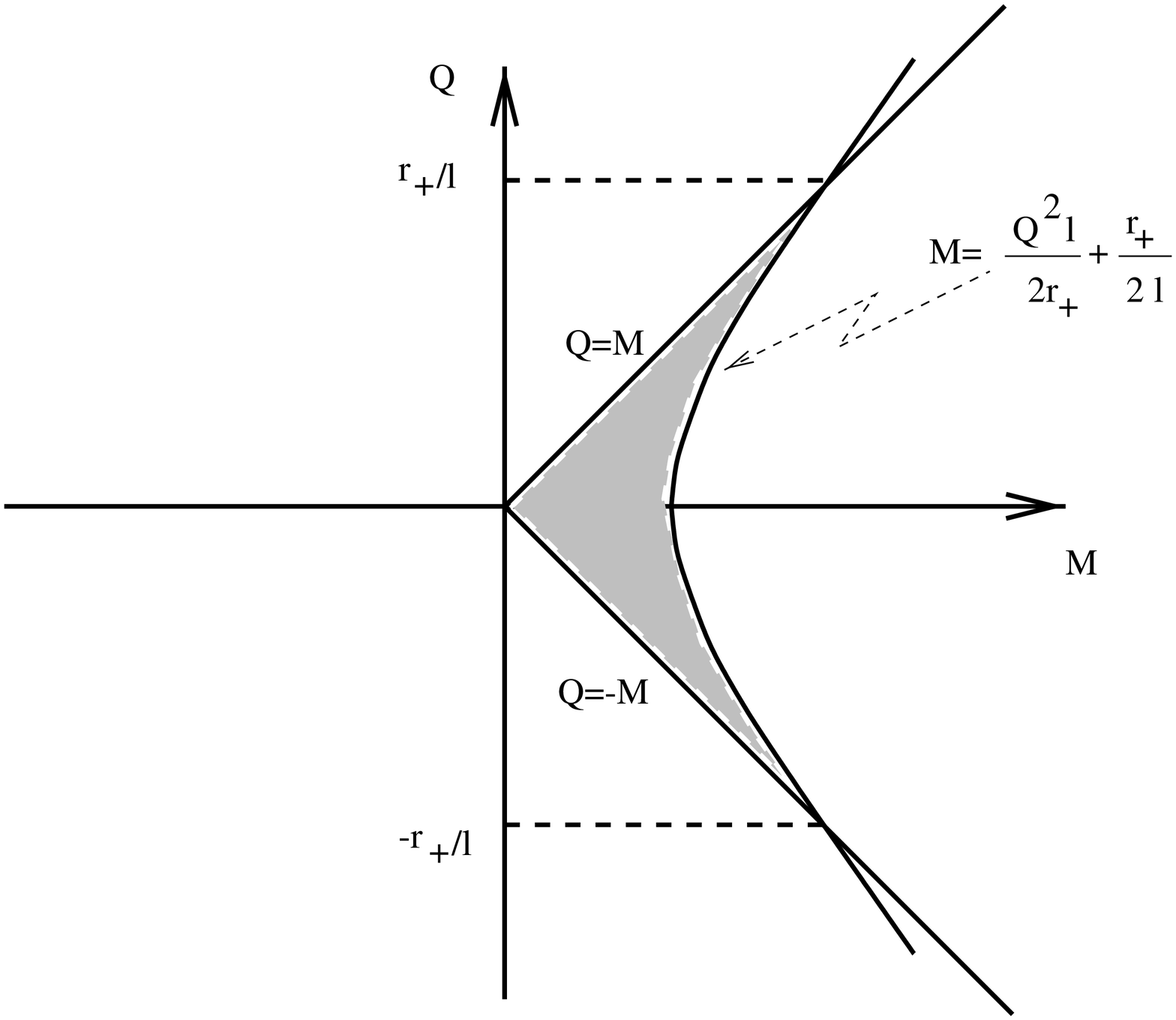}
\end{center}
\caption{Allowed space of observables for spherically symmetric gravity}
\label{ssg}
\end{figure}

Finally we examine in more detail the exact expression \req{eq: partition 3}
for the quantum partition function. In particular we will evaluate an
explicit expression for the volume of the allowed space of observables 
$\cal V$. Recall that we wish to restrict the values of $M$ and $Q$ so that 
there is always at least one positive and non-degenerate root to
$f(r,M,Q)=0$. Given the expression \req{eq: ssg horizons}, this requires:
\bea
M&>&0\\
M^2&>&Q^2
\label{eq: ssg bound1}
\eea
 Moreover, the outer boundary must lie exterior to the
outer horizon, so that 
\be
r_+> \lp^2(M+\sqrt{M^2-Q^2})
\ee
For any given value of charge $Q$, this puts an upper bound on the mass:
\be
M<{Q^2\lp^2\over 2r_+} + {r_+\over 2\lp^2}
\label{eq: ssg bound2}
\ee
The constraints \req{eq: ssg bound1} and \req{eq: ssg bound2} define the region
of observable space illustrated in Figure(\ref{ssg}). The volume of this space
can be readily obtained:
\bea
{\cal V} &=& \int^{r_+\over \lp^2}_{-{r_+\over \lp^2}}dQ
        \int^{{Q^2\lp^2\over 2r_+}+{r_+\over 2\lp^2}}_{|Q|}dM\\
  &=& {r_+^2/3 \lp^4}
\label{eq: ssg volume}
\eea

A numerical analysis of the partition function \req{eq: partition 3} for SSG will be treated
elsewhere.
\subsection{Dimensionally Reduced BTZ}
Starting with the Einstein action with cosmological constant in 2+1 
dimensions:
\be
I^{(3)} = {1\over 16\pi G^{(3)}}\int d^3x \sqrt{-g^{(3)}}
       (R(g^{(3)}) + \Lambda)
\ee
In 2+1 dimensions, the gravitational constant $G^{(3)}$ has dimensions of 
length. We now impose axial symmetry by  considering metrics of the 
form\footnote{In 2+1 dimensions, there is a generalized Birkhoff theorem
which states that all solutions have axial symmetry, and are stationary.}
\be
ds_{(3)}^2 = g_{\mu\nu}dx^\mu dx^\nu + \phi(x)^2(a d\theta + A_\mu dx^\mu)^2
\label{eq: axial ansatz}
\ee
where $a$ is an arbitrary constant with dimensions of length which, without
loss of generality we take to be proportional to the 2+1 dimensional
Planck length $a=8 G^{(3)}$. The one-form components $A_\mu$ are dimensionless. Unless the one-form $A=A_\mu dx^\mu$ is closed, the metric is not static
so that the field strength $F_{\mu\nu}= A_{\mu,\nu}-A_{\nu,\mu}$ is proportional to the angular momentum of the solution. With the above metric {\it ansatz}
the reduced action is that  of Jackiw-Teitelboim dilaton gravity 
coupled to an abelian gauge field:
\be
I^{(2)} =  \int d^2x \sqrt{-g}\left(\phi R(g) +\phi \Lambda
   -{1\over 4} \phi^3 F^{\mu\nu}F_{\mu\nu}\right)
\ee
This action is already of the generic form \req{eq: action 2} without
the need for further definitions. In particular, $G=1/2$, $l=\Lambda^{-1/2}$,
$V(\phi)= \phi$ and $W(\phi)= \phi^3$. Choosing $r=l\phi$ as the spatial
coordinate the general solution takes the form:
\be
ds^2 = -f(r,M,J)dt^2 + {1\over f(r,M,J)}dr^2
\ee
where
\be
f(r,M,J) = \left({r^2\over 2l^2} - Ml + {J^2 l^4\over 4 r^2}\right) 
\ee
As mentioned above, the abelian charge $J$ in this case is the angular momentum
of the black hole. For non-zero $J$ there are again two event horizons, at
\be
r_{o,i}=l\left(Ml\pm\sqrt{(Ml)^2 - (Jl)^2/2}\right)^{1\over 2}
\ee
where $r_o$ ($r_i$) is the outer (inner) horizon. The associated entropy is
\be
S=4\pi {r_o\over l} = 4\pi \phi(r_o) = {A\over 4 G^{(3)}}
\ee
where $A = 2\pi a \phi(r_o) = 16 \pi G^{(3)} \phi(r_o)$ is the invariant circumference
of the outer horizon, as calculated from \req{eq: axial ansatz}. The Bekenstein
Hawking entropy can also be calculated directly from \req{eq: Hawking temperature} to be 
\be
T_{BH} = {1\over 4\pi l^2} \left({r^2_o-r_i^2\over r_o}\right)
\ee 
In the semi-classical approximation, the mean energy of a black hole
in a box of fixed temperature and radius is:
\be
<E>= {r_+^2\over l^3}\left(1-\sqrt{1-{2\Mbar l^3\over r_+^2} + {\Jbar^2 l^6\over
                       2 r_+^4}}\right)
\ee
where $\Mbar$ and $\Jbar$ are the mean mass and angular momentum. Note that 
we have used the fact that the physical metric is $g_{\mu\nu}$ in this
case, so that $h(\phi_+)=j(\phi_+)$. The physical metric is not asymptotically
flat (it is in fact a metric of constant curvature) which accounts for 
the strange asymptotic behaviour of the mean energy as the box size
goes to infinity. One can
invert this relation to express the mass in terms of the mean energy:
\be
\Mbar= <E> - {<E>^2 l^3\over 2r_+^2} + {<J>^2l^3\over 4 r_+^2}
\ee
It is also straightforward to calculate the chemical potential. It is:
\be
\gamma = -{Jl^3\sqrt{1-r_o^2/r_+^2}\over 2r_o^2\sqrt{1- {J^2 l^6\over 
  2 r_o^2 r_+^2}}}
\ee
which approaches
\be
\gamma\to - {J l^3\over 2 r_o^2}
\ee
as $r_+\to \infty$.

Finally, we calculate the allowed volume $\cal V$ of the physical configuration
space. As in the case of spherically symmetric gravity we restrict $M>0$
and, in order that the horizons be non-degenerate $M>J/\sqrt{2}$. For the 
box size to be greater than the radius of the outer horizon, we also require,
\be
M<{J^2l^3\over 4 r_+^2 } + {r_+^2\over 2l^3}
\ee
The shape of the allowed configuration space is qualitatively as in Fig.(1),
but the slope of the straight lines is $1/\sqrt{2}$ and the parabola
has a different dependence on $r_+$. These conditions again put a bound
on the allowed range of $J^2$, namely:
$J^2< 2r_+^4/l^6$. The volume of the shaded region in this
case is:
\bea
{\cal V} &=& \int^{\sqrt{2}r_+^2/l^3}_{-\sqrt{2}r_+^2/l^3} dJ
   \int^{{J^2l^3\over 4 r_+^2 } + {r_+^2\over 2l^3}}_{J/\sqrt{2}} dM\\
    &=& {\sqrt{2}\over 3}{ r_+^4\over l^6}
\label{eq: JT volume}
\eea
\section{Conclusions}
We have calculated the Hamiltonian partition function for generic dilaton
gravity coupled to an Abelian gauge field. The class of theories considered
contains many specific charged black holes of physical interest. We  verified that our formalism gives the correct partition function in the saddle
point approximation for spherically symmetric gravity. We then used our
generic results to obtain  the partition function for a rotating BTZ black hole
confined to a box of fixed radius and temperature.
\par
In principle the partition function that we derived can be integrated exactly.
In practice, however, a numerical analysis is required in order to go beyond the semi-classical approximation. 
In a subsequent paper, we will do such a numerical analysis for specific theories, such as the BTZ black hole, in order to gain further information about
phase structure, specific heats, etc.
The {\it ansatz} that we used is, however,  only
 rigorous in the semi-classical
approximation. In particular, the integration measure, although motivated by
plausibility arguments, was not derived from the fundamental quantum theory, 
so it is  likely  that there are further quantum corrections that we have not
been able to encorporate. A detailed analysis of the possible quantum
corrections is currently in progress.
\section{Acknowledgements}
\par
 This work
was supported in part by the Natural Sciences and Engineering
Research
Council of Canada. G.K. would like to thank J. Gegenberg for helpful 
conversations.
  \par

\end{document}